\documentclass[11pt,onecolumn, peerreview,draftclsnofoot]{IEEEtran}
 
\usepackage[cmex10]{amsmath}
\usepackage{url}
\usepackage{graphicx}
\usepackage{color}
\usepackage{placeins}
\usepackage{float}
\usepackage{tabularx,colortbl}
\usepackage{multirow}
\usepackage{xfrac}
\usepackage{array}
\usepackage [autostyle, english = american]{csquotes}
\usepackage[table]{xcolor}    

\begin{document}

\title{Coexistence between Communications and Radar Systems - A Survey}

\IEEEoverridecommandlockouts

\author{\IEEEauthorblockN{Mina Labib*, Vuk Marojevic*, Anthony F. Martone\textsuperscript{\textdagger}, Jeffrey H. Reed*,  Amir I. Zaghloul*\textsuperscript{\textdagger} }
	\IEEEauthorblockA{\\ *Virginia Tech, Blacksburg, VA, USA\\
     \textsuperscript{\textdagger}US Army Research Laboratory, Adelphi, MD, USA}}
	
\maketitle

\begin{abstract}
Data traffic demand in cellular networks has been tremendously growing and has led to creating congested RF environments. Accordingly, innovative approaches for spectrum sharing have been proposed and implemented to accommodate several systems within the same frequency band. Spectrum sharing between radar and communications systems is one of the important research and development areas. In this paper, we present the fundamental spectrum sharing concepts and technologies, then we provide an updated and comprehensive survey of spectrum sharing techniques that have been developed to enable some of the wireless communications systems to coexist with radars in the same band.

\end{abstract}


\section{Introduction}

Billions of people rely on wireless infrastructures for communication and connectivity, to the extend that wireless communications became a necessary part of human life. Data traffic demand in cellular and wireless local area networks have been growing tremendously and has led to creating a congested Radio Frequency (RF) environment. Spectrum regulators adopted the fixed spectrum allocation policy, where bands were allocated to specific users or services. Since actual use of spectrum is, on average, way below its capacity, this leads to inefficient use of spectrum. Several studies have shown that,  apart from the spectrum used for wireless communications, a large portion of the spectrum is quite low-utilized \cite{cogRadioSurvey}. Consequently, the concept of spectrum sharing has gained lots of interest recently in order to help improving spectrum utilization. Spectrum sharing implies that two or more users (using different technologies) can share the spectrum and use it as needed and available, without creating harmful interference to one another.

Spectrum sharing is becoming possible because of advances in cognitive radio technology as well as more flexible spectrum regulations and incentives to share resources. Considering the United States for example, in June 2010, a \textit{Presidential Memorandum} titled \enquote{Unleashing the Wireless Broadband Revolution} was issued. The memorandum directed the National Telecommunications and Information Administration (NTIA) to collaborate with the Federal Communications Commission (FCC) to make a total of 500 MHz spectrum available over the following 10 years for wireless broadband systems usage, either on an exclusive license base or on a shared access base \cite{president1}. In June 2013, another \textit{Presidential Memorandum} titled \enquote{Expanding America's Leadership in Wireless Innovation} was issued, which urged to identify opportunities to share the spectrum that is currently allocated for exclusive use by Federal agencies. This memorandum explicitly requested to investigate the possibility of either relinquishing or sharing the spectrum of the 1695-1710 MHz band, the 1755-1850 MHz band, and the 5350-5470 and 5850-5925 MHz bands \cite{president2}. Consequently, the Department of Defense (DoD) issued a \enquote{A Call To Action,} where one of the objectives was to \enquote{accelerate the fielding of technologies that enable spectrum sharing and improve access opportunities,} and one of the goals was to \enquote{sharpen responsiveness to ongoing spectrum regulatory and policy changes} \cite{dod1}. In January 2015, the FCC completed an auction to license the 1695-1710 MHz, 1755-1780 MHz, and 2155-2180 MHz frequency bands to wireless operators. These bands are collectively called the AWS-3 band (Advanced Wireless Service). Most of the Federal systems will relocate to new bands, while few systems remain and share the spectrum with the awarded wireless operators. The auction generated more than 42 Billion dollars in net profits. In March 2015, the DoD Chief Information Officer issued a brief to the DoD UAS (Unmanned Aerial Systems) summit, where he indicated that spectrum access is being challenged by emerging commercial market needs and that DoD is taking deliberate actions to advocate spectrum sharing techniques \cite{dod2}. In April 2015, NTIA announced a 12-month plan to cooperate with FCC and other stakeholders to study and develop sharing options that accommodate new applications and devices with the reserved 195 MHz of the 5 GHz band. Currently, these 195 MHz (5350-5470 MHz and 5850-5925 MHz) are mainly used by radar systems, and the major user of these bands is the DoD. The DoD uses this for different radar systems, which are mainly Range and Tracking Radars, Tactical Anti-Air Warfare Radar Systems, Navigation Radars, and Weather Radars \cite{deptCommerce1}. 

Radar systems are the main consumers of the frequency bands that are currently being considered for sharing in the US, such as the 5150-5925 MHz, which is called Unlicensed National Information Infrastructure (U-NII) band, and the 3550-3700 MHz band, which is called Citizen Broadband Radio Service (CBRS). Accordingly, spectrum sharing between communications and radar systems has become one of the important research and development areas. When a radar system shares the spectrum with a communications system, the interference caused to the radar can impede its correct functioning. This is so because radars were not designed to coexist with communications. Neither were communications systems designed to operate with radars. Nevertheless, communications systems  are easier to modify than legacy radars because of the much shorter development and deployement cycles. 

The authors in \cite{radar_sharing_survey} provide an overview of some of the techniques that have been proposed for sharing the spectrum between radar and wireless communications systems. In this paper, we present the fundamental spectrum sharing concepts and technologies, then we provide an updated and more comprehensive survey of spectrum sharing techniques that have been developed to enable some of the wireless communications systems to coexist within the same band of radar systems. The rest of the paper is organized as follows: Section II provides an analysis of the different radar systems that exist within the shared spectrum. Section III provides a general overview of the spectrum sharing techniques. Section IV present our survey of the proposed spectrum sharing mechanisms for radar and communications systems. We provide our conclusions in Section V.

\section{The Shared Spectrum: Radar Systems and Regulations}

There are several frequency bands that are utilized by radar systems and they are currently considered for spectrum sharing in the US, such as U-NII and the CBRS bands. In order to emphasize the need for developing spectrum sharing techniques for the coexistence between communications and radar systems, this section presents the different radar types within these two bands, as well as the current regulations governing these bands. 

For the CBRS band, the frequency range 3550-3650 MHz is allocated on a primary basis for federal use of both the Radiolocation Service (RLS) and the Aeronautical Radionavigation Service (ARNS) (ground-based). The RLS radar is used by the military for the purpose of radiolocation, whereas the ARNS radar is used for the safe operation of military aircrafts \cite{fcc3.5}. The authors in \cite{Munwar1} analyze the regulations governing this band.  

For the 5 GHz band, there are several radar systems operating within this band: 

\begin{itemize}
	\item Radiolocation service for federal operation: The frequency range 5250 - 5950 MHz is allocated to radiolocation services for federal operation on a primary basis. The frequency range 5470-5650 MHz is allocated to radiolocation services for non-federal operation on a primary basis as well. These types of radars perform variety of functions such as \cite{itu1}:
	\begin{itemize}
		\item Tracking space launch vehicles during the developmental and operational testing phases,
		\item Sea and air surveillance,
		\item Environmental measurements,
		\item National defense.
	\end{itemize}
	
	\item Space Research Services: Band U-NII-2A and the frequency segment 5470-5570 MHz are both allocated on a primary basis to space research services for federal operation and on a secondary basis to the space research services for non-federal operation. 
	\item Aeronautical Radio Navigation Radars: These radars use the frequency range 5350-5460 MHz (U-NII-2B). These types of radars are used for:
	\begin{itemize}
		\item airborne weather avoidance,
		\item windshear detection,
		\item safety service for flights.
	\end{itemize}
	
	\item Maritime Radio Navigation Radars: The frequency range 5350-5460 MHz is allocated on a primary basis to the maritime radio navigation radars for both federal and non-federal operations. These types of radars are used for the safety of ships. 
	
	\item Terminal Doppler Weather Radars (TDWR): The Federal Aviation Administration (FAA) uses the frequency range 5600-5650 MHz for TDWR to improve the safety of operations at several major airports (45 airports) as these radars provide quantitative measurements for identifying weather hazards.

\end{itemize}

Furthermore, the DoD uses the 5 GHz frequency band for different military radar systems. The DoD radars can either operate on a fixed frequency or employ frequency hopping techniques. In the past, these radars have operated on or near military installations. But, in support of homeland security, these radars may need to be deployed in urban areas, where U-NII devices (commercial and industrial) will be heavily used, as is the case of today's 5 GHz Wi-Fi systems \cite{deptCommerce1}. So, the interference issue between radar systems and the U-NII devices needs to be addressed. These radar systems are used by different DoD agencies: the U.S. Army, the U.S. Navy and U.S. Coast Guard, and the U.S. Air Force. The major DoD radar systems operate within the 5 GHz band can be summarized as follows \cite{deptCommerce1}:

\begin{itemize}
	\item Range and Tracking Radars: These radars operate in 5400-5900 MHz frequency band. During the development and testing phases of non-cooperative targets (such as rockets and missiles), these radars are used to provide accurate tracking data of these targets and to maintain range safety. 
	\item Weather Radars: These radars are used to determine the structure of weather hazards by analyzing the location and the motions of several types of precipitations (i.e., rain, snow, hail, etc.). 
	\item Shipboard Navigation Radars: These radars operate in the 5450-5825 MHz frequency range. The main function of this radar type is detecting anti-ship missiles and low flying aircraft. These radars are also used for navigation and for general surface search tasks. 
	\item Airborne Sense and Avoid Radars: The U.S. Air Force has been developing these types of radars to comply with the safety concerns of the FAA for Unmanned Aerial Vehicle (UAV) operations. The purpose of this radar is to avoid collision of the UAV by detecting and tracking other nearby aircrafts. The 5350-5460 MHZ frequency range is the one that is currently considered for these radars, whereas the frequency range 5150-5250 MHz is being considered as an alternative. 

\end{itemize}

\begin{figure}[h]
	\centering
	\begin{center}
		\includegraphics[width=5in]{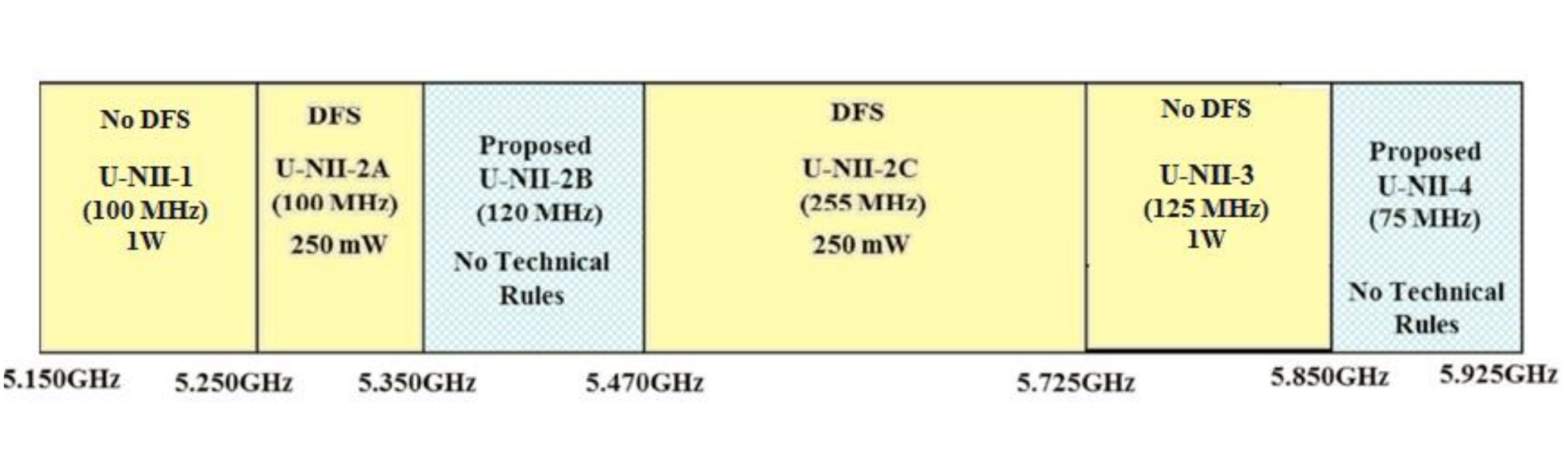} 
		\vspace{-0.25in}
		\caption{The 5 GHz regulations in the United States.}
		\label{5ghz_US}
	\end{center}
	\vspace{-0.25in}
\end{figure}

The 5 GHz band falls under the FCC Part-15 regulations. The current spectrum regulations for the 5 GHz band in the US are depicted in Figure \ref{5ghz_US} \cite{fcc1}, which identifies the maximum transmit power level allowed for devices in each sub-band.

Any device operating in the U-NII-2A or the U-NII-2C bands must employ Transmit Power Control (TPS) and Dynamic Frequency Selection (DFS). DFS is a mechanism that is specifically designed to avoid causing interference to non-IMT (International Mobile Telecommunications) systems, such as radars. The DFS requirements in the United States can be summarized as follows:
\begin{itemize}
	\item Sensing bandwidth: The device must sense radar signals in 100\% of its occupied BW. 
	\item Channel availability check time: The device must check the channel for sixty seconds before using it. 
	\item In-service monitoring: The device must continuously monitor the channel during operation and must vacate the channel within ten seconds (called Channel Move Time) once the radar system start transmitting. During these 10 s, the device is only allowed 200 ms for normal transmission.  
	\item Detection threshold: It is the received power levels when averaged over one microsecond referenced to a 0 dBi antenna. Specifically:
	\begin{itemize}
		\item -62 dBm: For devices with maximum EIRP (Effective Isotropic Radiated Power) less than 200 mW (23 dBm) and an EIRP spectral density of less than 10 dBm/MHz (10 mW/MHz)
		\item -64 dBm: For devices that do not meet the above requirement for relaxed sensing detection.  
	\end{itemize}
	\item Detecting radar: once the radar has been detected, the operating channel must be vacated. The device must not utilize the channel for thirty minutes, which is called Non-Occupancy Period. 
	
\end{itemize}

The FCC has identified six different radar waveforms that need be used for testing the U-NII devices that operate in this band. Table \ref{tab:radarwaveforms} provides the five test waveforms as defined by FCC \cite{deptCommerce1}. The sixth waveform is for frequency hoping radar, with pulse width 1 $\mu$s, pulse repetition interval of 333 $\mu$s, and 9 pulses per frequency hop, hoping sequence length of 300 ms and hoping rate of 333 Hz. The minimum detection probability of this test is 70\%.  For all of these test waveforms, the minimum number of trials to ensure that the device is passing the test is 30 trials. 

\begin{table}
	\centering
	\caption{Radar test waveforms for DFS Algorithm}
	\label{tab:radarwaveforms}
	\vspace{0.1in}
	\begin{tabular}{| p{1.5cm} | p{1.5cm} |p{2cm} |p{2cm} |p{1.5cm} |p{1.7cm} |p{2.5cm} |}
		\hline
		\textbf{Radar Type} & \textbf{Pulse Width($\mu$s)} & \textbf{Pulse Repetition Interval($\mu$s)} & \textbf{Number of Pulses} & \textbf{Number of bursts} & \textbf{Minimum Detection Probability} & \textbf{Comments} \\ \hline
		1 & 1 & 1428 & 18 & N/A & 60\% & Short Pulse Radar \\ \hline
		2 & 1-5 &  150-230& 23-29 & N/A & 60\% &  Short Pulse Radar \\ \hline
		3 & 6-10 & 200-500 & 16-18 & N/A & 60\% &Short Pulse Radar \\ \hline
		4 & 11-20 & 200-500 & 12-16 & N/A & 60\% &Short Pulse Radar \\ \hline
		5 & 50-100 & 1000-2000 & 1-3 & 8-20 &80\% & Long Pulse Radar, with chirp width 5-20 MHz \\ \hline

	\end{tabular}
	
\end{table}

\section{Spectrum Sharing Overview}

Dynamic spectrum sharing involves two main tasks: spectrum awareness and dynamic spectrum access (DSA). 

\subsection{Spectrum Awareness}
\begin{figure}[h]
	\centering
	\begin{center}
		\includegraphics[width=5in]{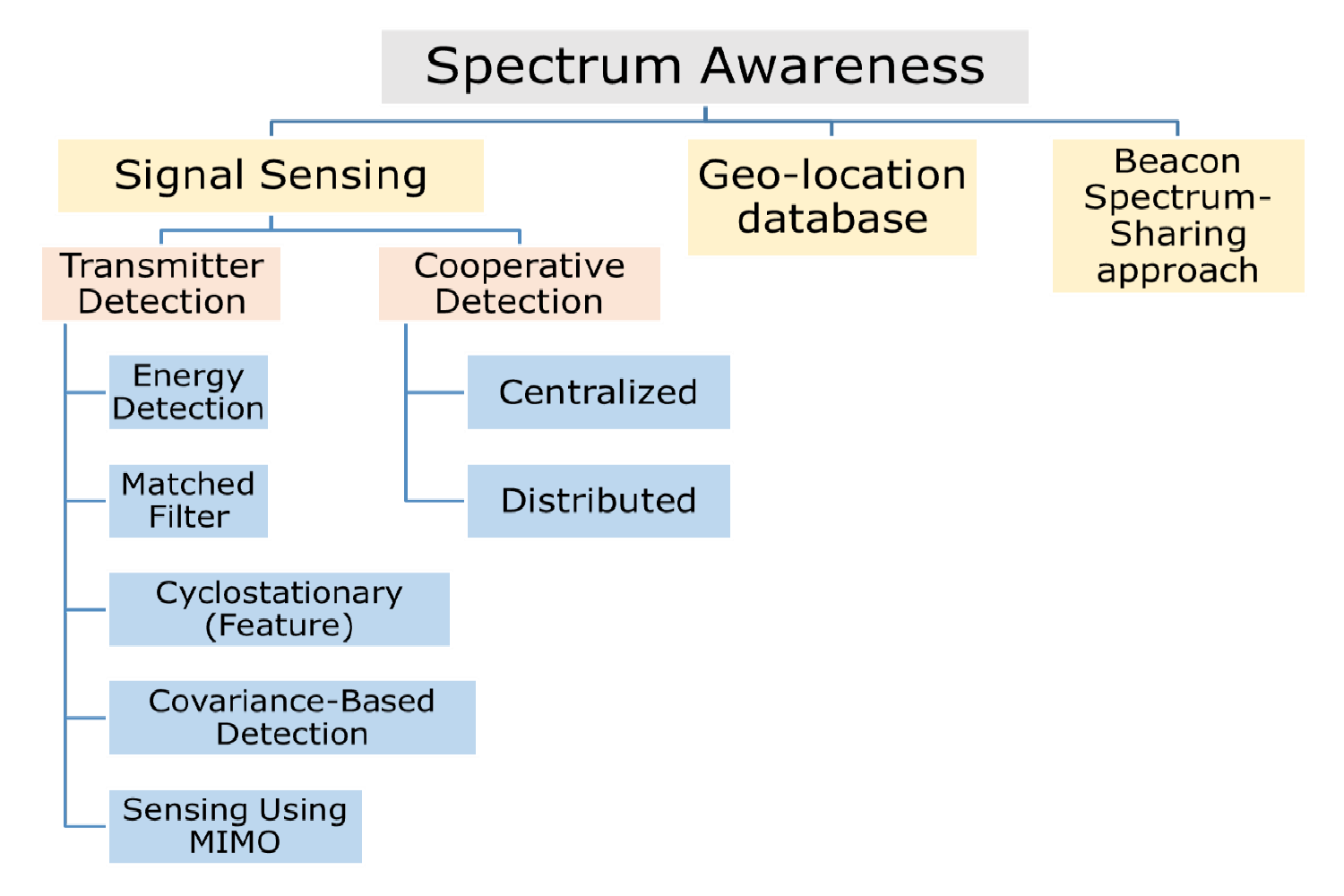} 
		\caption{Common spectrum awareness techniques.}
		\label{spectrumAware}
	\end{center}
\end{figure}

Spectrum awareness refers to the way users capture information about the RF environment in order to be aware about other users using the spectrum. Figure \ref{spectrumAware} illustrates the common spectrum awareness techniques. 

Signal sensing  refers to tuning to the band of interest, receiving and processing I/Q samples to assess if there are detectable signals in the spectrum. This can be subdivided into two types: cooperative detection and transmitter detection. Cooperative detection refers to the case where there are several sensors that are geographically spread and share the spectrum measurements. If the information collected from different sensors is processed at a central unit, then it is called centralized detection. If the information collected from different sensors is shared among them, but each device makes its own decision, then it is called distributed detection \cite{deptCommerce1}. 
For transmitter detection, the following techniques have been proposed \cite{cogRadioSurvey}:
\begin{itemize}
	\item Energy Detection (ED): Compares the energy received in a given band with a pre-defined threshold. This algorithm does not require knowing the nature of the transmitted signal, but does not perform well under low SNR (Signal-to-Noise Ratio) conditions.
	\item  Matched Filter: Correlates the received signal with the known signal. This algorithm requires knowing the characteristics of the transmitted signal and performs well in low SNR conditions. 
	\item Cyclostationary Feature Detection: Correlates the received modulated signals with either sine-wave carriers, pulse trains, or cyclic prefixes. And that is based on knowing certain features of the received signal. The signal detection is performed by analyzing the spectral correlation function. 
	\item Covariance-based detection (CBD): The detection is performed by comparing the correlations between the received signal and the non-zero lags with the correlations between the received signal and the zero lag. 
	\item Sensing using MIMO (Multiple-Input Multiple-Output) systems: when the device is equipped with multiple antennas, it can use Eigenvalue-based detection (EBD) for spectrum sensing. 
\end{itemize}

For spectrum sensing, it is important that the devices monitor the channel initially and periodically as well. Some of the important parameters for spectrum sensing algorithms are the detection threshold level, how often and how long the channel needs to be monitored.

Another method for spectrum awareness is using a Geo-location database, which is a useful for systems with fixed locations, as in the case of fixed radar systems (such as TDWR systems). In this method, the device is equipped with Global Position System (GPS) and it has access to a database that contains the locations of systems that has the right to use the spectrum in different geographical areas \cite{deptCommerce1}. 
In beacon spectrum sharing approaches, the device that wishes to access the spectrum as secondary user must have the ability to receive certain control signals sent by the incumbent system during the period that spectrum sharing is allowed. The device should not start transmitting without receiving this beacon signal and it should stop once it stops receiving it \cite{deptCommerce1}.  

\subsection{Dynamic Spectrum Access}
There are two main models for DSA: the opportunistic spectrum access (OSA) and the concurrent spectrum access (CSA). For OSA, one of the users has the priority access to the spectrum. This type of user is called Primary User (PU). Any other user sharing the spectrum with the PU is called Secondary User (SU).
the SU cannot access the spectrum as long as the PU is using it. The  SU will use spectrum sensing or query a database to determine the times that the primary user is not using the channel in order to access it opportunistically. A good example for such approach is the DFS algorithm mandated in U-NII-2A and U-NII-2C bands. For CSA, different users are allowed to share the spectrum together as long as it is done in a fair manner and as long as the interference generated at each receiver is below a certain threshold. In the next sections, we will examine the spectrum sharing techniques proposed for the coexistence between communications and radar systems.

\section{Spectrum sharing between communications and radar Systems}

DARPA (Defense Advanced Research Projects Agency) has launched the SSPARC (Shared Spectrum Access for Radar and Communications) program to promote research and development for spectrum sharing between communications and radar systems \cite{SSPARC}. In order to develop an effective spectrum sharing algorithm, it is important to first analyze the effect that each system has on the other. 

For analyzing the effect of communications system on radar performance, the authors in \cite{khawar2014mathematical} provide a mathematical model for the interference generated from a cellular system on radars in the 3.5 GHz band. They prove that the interference generated from cellular systems in correlated shadow fading channel has a log-normal distribution and provide the bounds on the radar's probability of detection under cellular interference when there is no constraint on the transmitted power level of the cellular system. The paper shows that such interference can considerably degrade the performance of a radar system in terms of accurate detection of targets. The work in \cite{6903972_Results_SS} uses simulations to show that communications system interference can significantly degrade the radar performance. The authors in \cite{inBandOFDMIntrf} confirm the same conclusion after examining the impact of in-band OFDM interference on radars. In \cite{5582021_Radar_LTE_2300}, the authors perform system level analysis for the coexistence between a Time Division Long Term Evolution (TD-LTE) system and a radar system in the band 2300-2400 MHz and study the effect on each other under different isolation distance and frequency spacing. 

For analyzing the effect of radar on communications system performance, the authors in \cite{7485064_EffectOfRadarInterf} provide an analytical model to study the effect of unaltered radar signal on the bit error rate of a communications system. The authors in \cite{6192906_Radar_LTE_2300} have analyzed the bit error rate of TD-LTE in the frequency band 2300-2400 MHz with the presence of radar signal interference. In \cite{clencylte35}, a system level analysis was performed for quantifying the impact of shipborne radar on an outdoor LTE base station (eNodeB) in the 3.5 GHz band. The results show that radar signal can decrease the SINR (signal-to-interference-plus-noise ratio) at the LTE eNodeB, but the SINR level improves in between radar pulses. The results also show that the exclusion zones set by the NTIA to protect the performance of the commercial communications systems when sharing the spectrum at the 3.5 GHz are too conservative. The same conclusion is confirmed in \cite{7462190_Reed_3.5GHz}, where a field experiment was performed in St. Inigoes, Maryland for the coexistence between TD-LTE system and a land-based AN/SPN-43C radar in the 3.5 GHz band. 
 
These research contributions illustrate that when a radar system shares the spectrum with one of the communications systems, it is more vulnerable to interference because of the radar system's sensitivity. Furthermore, LTE systems in particular deploy several interference management mechanisms to reduce the effect of interference on the performance. For the downlink, the eNodeB can deploy a Frequency-Selective Scheduler (FSS), which uses the Channel Quality Indicator (CQI) reporting from the UE and perform a frequency domain scheduling and avoid assigning Physical Resource Blocks (PRBs) that have excessive interference or fading to that UE. Moreover, LTE can deploy interference shaping to minimize the inter-cell interference, where a low-loaded eNodeB does not rapidly change its frequency-domain allocations in order to improve the quality of CQI reporting for the UEs served by a neighbor high-loaded eNodeB \cite{LTESmallCell}. For the uplink, LTE uses both open-loop and closed-loop power control with interference awareness, where the eNodeB takes into account the neighbor cells when adjusting the uplink power transmission for the UEs, thus reducing the inter-cell interference and improving the overall uplink data rates \cite{nokia_sch}. 

Several research campaigns have investigated developing spectrum sharing techniques for the coexistence between radar and communications systems. As shown in \cite{radar_sharing_survey}, the approaches for radar and communications systems can be broadly classified into three broad categories:

\begin{itemize}
	\item Cognitive communications system: These communications systems are aware of the RF environment and can dynamically adapt to avoid harmful interference to the radar systems.
	\item Cognitive radars: The radar is cognitive and responsible for not affecting the performance of  the communications system.
	\item Joint Cognition: Both the radar and the communications system collaborate to avoid creating harmful interference to each other.  
\end{itemize}

\subsection{Cognitive Communications Systems} 

The term \enquote{cognitive radio} refers to a system that has the ability to sense the surrounding RF environment, to make short term predictions, to learn how to operate in a given environment, and to dynamically adjust its transmitting parameters. These capabilities are used by a communications system that acts as a secondary user when sharing the spectrum with radar systems. The first example of such category is the Dynamic Frequency Selection (DSF) mandated by the FCC for devices operating in the 5 GHz band. A very important aspect of the DSF is spectrum sensing. Accordingly, several research efforts have been done to improve the sensing capabilities of the communications system, such as employing cooperative sensing using multiple sensors \cite{4557030_CoperativeSensing}  \cite{4917628_CopertaiveSensing2}. 
The authors in \cite{7500412_RadioMap} investigate using a Radio Environment Map (REM) to enhance the spectrum awareness for communications systems sharing the frequency band with radars. In \cite{7537114_MIMO_Com_5GH}, the authors explore using multiple transmit antennas at the communications system side for sharing the spectrum with radars in the 5 GHz band, while still employing DFS and adhering to the regulatory constraints of the maximum transmit power. 

The authors in \cite{rotatingRadar1} introduce a spectrum sharing algorithm to address the coexistence issue between an OFDM communications system and a rotating radar. One of the main requirements for this algorithm is that the base stations, which is the secondary user, needs to be able to communicate with its mobile terminals over a different frequency band in addition to the one shared with the radar system. Another constraint is that the base station must not generate harmful interference to the radar systems, determined by maximum interference-to-noise ratio threshold at the radar. In other words, the detection capabilities of the radar must not be comprised. The basic concept is that the communications system can utilize the period when the antenna beam of the rotating radar is not within the range of the communications system transmission. They show that communications systems can achieve a significant increase in the downlink throughput if they can tolerate the interruptions due to radar rotations. The authors in \cite{7131242_AirTrafficRadar} investigate and validate the feasibility of adjacent-channel coexistence of an LTE system and an air traffic control radar (with a rotating main beam) in the L-band through system level simulation. 
 
The authors in \cite{5872307_Wimax} investigate the coexistence between WiMax (Worldwide Interoperability for Microwave Access) networks and nearby groud-based radar systems that operate in adjacent frequency channel within the S-band. They propose four different interference mitigation techniques for WiMax using either spatial, spectral, temporal or system-level modifications domains.

\subsection{Cognitive Radars} 

The concept of cognitive radar was first introduced in \cite{1593335_haykin}, cognitive radar techniques have shown the ability  to improve the overall performance of the radar \cite{martone2014cognitive}. With cognitive techniques and with MIMO technology, radar systems could become more resilient by implementing interference mitigation algorithms. When the radar acts as the secondary user, it will need to perform spectrum sensing as shown in \cite{6956678_CogRadar_OverviewSpecSending} and \cite{7410950_CogRadar_SpecSensing}. The approaches for spectrum sharing while using cognitive radars can be categorized into two sub-categories: waveform shaping and waveform design. 

\subsubsection{Waveform Shaping} 
When a radar employs a MIMO system, it becomes capable of reshaping the radar waveform. In \cite{deng2013}, a novel algorithm has been proposed for coherent MIMO radar that can minimize the interference generated by communications systems signals coming from any direction; the main lobe or the side lobes direction. In this approach the radar transmits and receives coherent orthogonal phase-coded waveforms from each of the antenna elements. The work in \cite{6933960_AdaptiveRadarBeamforming} provides the mathematical proof for this approach and the conditions for the MIMO radar main lobe interference cancellation. The authors in \cite{6636787_Radar_Precoder} provide a practical precoding approach for the coexistence between MIMO radar and multiple communications transmitter-receivers pairs, where the radar transmitter applies a zero-forcing precoder that eliminates the interference at the communications system side. This approach requires perfect knowledge of the interference channel matrix at the radar side. So they propose a channel estimation algorithm that requires the multiple communications transmitters to coordinate in transmiting training symbols. In \cite{sodagari2012}, the authors propose an algorithm for spectrum sharing between a MIMO radar and a communications system by projecting the radar waveform into the null space of the interference channel matrix between the two systems. The authors in \cite{6817773_SS_SpatialApprch} extend that work from one communications system into multiple ones and provide a new projection algorithm for this case. The authors in \cite{6956861_ImpactOfChannel} extend the work in \cite{sodagari2012} for a maritime MIMO radar, where they consider the variations in the channel due to the motion of the ship.

\subsubsection{Waveform Design} 
Since cognitive radars have the ability to modify their transmitted waveform based on the preceding returns, the authors in \cite{6810549_radarwaveformDesign} investigate a radar waveform design for the coexistence  with a friendly communications system and propose using SNR based waterfilling technique that minimizes the interference at the communications system. In \cite{radarwaveAmuru}, the authors propose information theoretic waveform design for MIMO radars in the frequency domain to reduce the amount of interference produced to a coexisting communications system. The authors in \cite{awais2014} design a finite alphabet binary phase shift keying (BPSK) waveform for stationary and moving maritime MIMO radar in a way that does not generate interference to the communications systems operating in the same band. In \cite{7472362_MutualInfo_Joint}, the authors introduce a Mutual Information (MI) based criterion for OFDM radar waveform optimization in a joint radar and multiple base stations setup.

\subsection{Joint Cognition} 
In recent year, joint radar and communications system spectrum sharing techniques have gained a lot of interest because of the potential to enhance the spectrum sharing performance due to the additional degrees of freedom. In \cite{bhat2012}, the authors introduce the concept of bandwidth sharing between multimodal radar and a communications system. A multimodal radar has the ability to vary its bandwidth (and accordingly its resolution) based on its current needs; this allows the communications system to utilize the remaining bandwidth. The priority for radar is determined using fuzzy logic and the priority for the communications system was assumed. The authors in \cite{martone3} point out the challenge of this approach given the congestion of the RF spectrum, and hence develop novel spectrum sharing techniques for adaptable radars, where the radar selects, using multi-objective optimization, the optimal sub-band in a highly congested frequency band to both maximize the SINR and minimize the range resolution cell size. The authors in \cite{6586097_ofdmWaveform} propose to design the OFDM waveform for both the multimodal radar and a communications system by assigning OFDM sub-carriers based on maximizing the radar detection performance and the communications system channel capacity. The authors in \cite{7131098_Joint} propose a joint design and operation between radar and a communications system based on an transmit and receive optimization procedures to maximum forward channel SNR while simultaneously minimizing co-channel interference. The authors in \cite{7470514_OptimumCoDesign} present an algorithm for the co-design of MIMO radar and MIMO communications system for spectrum sharing based on a cooperative approach to mitigate mutual interference. The work in \cite{7485158_MIMO_RadarClutter} extends this cooperative approach for a radar system operating in the presence of clutter. The work in \cite{7472289_JointDesign} extends it further for a scenario where all radar targets fall in different range bins.

\section{Conclusion}
Spectrum sharing between radar and communications systems is one of the important solutions proposed to help with the RF congestion problem. The impact this technology may have is proportional to the amount of spectrum affected, and radar systems consume huge amount of spectrum, while operating. In this paper, we have presented the different radar systems within the 5 GHz along with the spectrum regulations in this band. We have presented the basic concepts of spectrum sharing and performed an updated survey of the spectrum sharing contributions for the coexistence between communications and radar systems. We conclude that spectrum sharing between radar and communications systems is definitely feasible, but a lot of research is still required in order to ensure that the radar performance will not be affected, while improving the spectrum efficiency for communications.


\begin{thebibliography}{1}
	
	
\bibitem{cogRadioSurvey}
Y.~C. Liang, K.~C. Chen, G.~Li, and P.~Mahonen, ``{Cognitive Radio Networking
	and Communications: An Overview},'' in \emph{{IEEE Transactions on Vehicular
		Technology,}}, \textbf{60}, 7, September 2011, pp. 3386-3407.

\bibitem{president1}
{Office of the Press Secretary}, ``{Presidential Memorandum: Unleashing The
	Wireless Broadband Revolution},'' {The White House}, {Press Release}, June
2010. [Online]. Available:
\url{https://www.whitehouse.gov/the-press-office/presidential-memorandum-unleashing-wireless-broadband-revolution}

\bibitem{president2}
------, ``{Presidential Memorandum: expanding America's Leadership in Wireless
	Innovation},'' {The White House}, {Press Release}, June 2013. [Online].
Available:
\url{https://www.whitehouse.gov/the-press-office/2013/06/14/presidential-memorandum-expanding-americas-leadership-wireless-innovatio}

\bibitem{dod1}
{Deputy Secretary Of Defense}, ``{Electromagnetic Spectrum Stagey},''
{Department of Defense}, {A call To action}, September 2013.

\bibitem{dod2}
{F. D. Moorefield, DoD Chief Information Officer}, ``{Assured Dynamic Spectrum
	Access, Evolving toward Revolutionary Change: DoD Prospective},'' {Department
	of Defense}, {Brief to DoD UAS Summit}, March 2015.

\bibitem{deptCommerce1}
{Rebecca Blank and Lawrence E. Strickling}, ``{Evalution of the 5350-5470 MHz
	and 5850-5925 MHz Bands Pursuant to Section 6406(b) of The Middle Class Tax
	Relief And Job Creation Act of 2012},'' {Department of Commerce}, {Technical
	Report}, January 2013.

\bibitem{radar_sharing_survey}
H.~T. Hayvaci and B.~Tavli, ``{Spectrum Sharing in Radar and Wireless
	Communication Systems: A Review},'' International Conference on
Electromagnetics in Advanced Applications (ICEAA), August 2014, pp. 810-813.

\bibitem{fcc3.5}
{FCC}, ``{Amendment of the Commission’s Rules with Regard to Commercial
	Operations in the 3550-3650 MHz Band},'' {Federal Communications Commission},
{Report And Order And Second Further Notice of Proposed Rulemaking}, April
2015.

\bibitem{Munwar1}
M.~M. Sohul, M.~Yao, T.~Yang, and J.~H. Reed, ``Spectrum Access System for the
Citizen Broadband Radio Service,'' in \emph{IEEE Communications Magazine},
\textbf{53}, 7, July 2015, pp. 18-25.

\bibitem{itu1}
{ITU}, ``{Characteristics of and Protection Criteria for Sharing Studies for
	Radiolocation (except Ground Based Meteorological Radars) and Aeronautical
	Radionavigation Radars Operating in the Frequency Bands between 5 250 and 5
	850 MHz},'' {International Telecommunication Union}, {Recommendation ITU-R
	M.1638-1, M-Series}, January 2015.

\bibitem{fcc1}
{FCC}, ``{Revision of Part 15 of the Commission’s Rules to Permit Unlicensed
	National Information Infrastructure (U-NII) Devices in the 5 GHz Band},''
{Federal Communications Commission}, {First Report and Order}, April 2014.

\bibitem{SSPARC}
{Joseph B. Evans}, ``{Shared Spectrum Access for Radar and Communications
	(SSPARC)},'' {DARPA}, {Press Release}. [Online]. Available:
\url{http://www.darpa.mil/program/shared-spectrum-access-for-radar-and-communications}

\bibitem{khawar2014mathematical}
A.~Khawar, A.~Abdelhadi, and T.~C. Clancy, ``{A Mathematical Analysis of
	Cellular Interference on the Performance of S-Band Military Radar Systems},''
IEEE Wireless Telecommunications Symposium (WTS), April 2014, pp. 1-8.

\bibitem{6903972_Results_SS}
M.~R. Bell, N.~Devroye, D.~Erricolo, T.~Koduri, S.~Rao, and D.~Tuninetti,
``Results on Spectrum Sharing Between a Radar and a Communications System,''
International Conference on Electromagnetics in Advanced
	Applications (ICEAA), August 2014, pp. 826-829.

\bibitem{inBandOFDMIntrf}
B.~D. Cordill, S.~A. Seguin, and L.~Cohen, ``{Electromagnetic Interference to
	Radar Receivers due to In-Band OFDM Communications Systems},''
	IEEE International Symposium on Electromagnetic Compatibility, August 2013,
pp. 72-75.

\bibitem{5582021_Radar_LTE_2300}
W.~Liu, J.~Fang, H.~Tan, B.~Huang, and W.~Wang, ``{Coexistence Studies for
	TD-LTE with Radar System in the Band 2300-2400 MHz},'' International Conference on Communications, Circuits and Systems (ICCCAS), July 2010, pp. 49-53.

\bibitem{7485064_EffectOfRadarInterf}
N.~Nartasilpa, D.~Tuninetti, N.~Devroye, and D.~Erricolo, ``{Let's Share
	CommRad: Effect of Radar Interference on an Uncoded Data Communication
	system},'' IEEE Radar Conference (RadarCon), May 2016, pp. 1-5.

\bibitem{6192906_Radar_LTE_2300}
J.~Han, B.~Wang, W.~Wang, Y.~Zhang, and W.~Xia, ``{Analysis for the BER of LTE
	System with the Interference from Radar},'' IET International
	Conference on Communication Technology and Application (ICCTA 2011), October 2011, pp. 452-456.

\bibitem{clencylte35}
M.~Ghorbanzadeh, E.~Visotsky, P.~Moorut, W.~Yang, and C.~Clancy, ``{Radar
	In-Band and Out-Of-Band Interference into LTE Macro and Small Cell Uplinks in
	the 3.5 GHz Band},'' IEEE Wireless Communications and Networking Conference (WCNC), March 2015, pp. 1829-1834.

\bibitem{7462190_Reed_3.5GHz}
J.~H. Reed, A.~W. Clegg, A.~V. Padaki, T.~Yang, R.~Nealy, C.~Dietrich, C.~R.
Anderson, and D.~M. Mearns, ``{On the Co-Existence of TD-LTE and Radar over
	3.5 GHz Band: An Experimental Study},'' in \emph{IEEE Wireless Communications Letters,} \textbf{5}, 4, August 2016, pp. 368-371.

\bibitem{LTESmallCell}
R.~Agrawal, A.~Bedekar, H.~Holma, S.~Kalyanasundaram, K.~Pedersen, and
B.~Soret, ``Small cell interference management,'' in  H. Holma, A. Toskala, J. Reunanen, \emph{LTE Small Cell
	Optimization: 3GPP Evolution to Release 13, First Edition}, West Sussex, United Kingdom, John Wiley \& Sons Ltd, pp. 91-120, 2016.

\bibitem{nokia_sch}
{Nokia Solution and Networks}, ``{Smart Scheduler},'' 2015.

\bibitem{4557030_CoperativeSensing}
L.~S. Wang, J.~P. Mcgeehan, C.~Williams, and A.~Doufexi, ``Application of
Cooperative Sensing in Radar-Communications Coexistence,'' in \emph{IET Communications,} \textbf{2}, 6,
July 2008, pp. 856-868.

\bibitem{4917628_CopertaiveSensing2}
L.~Wang, A.~Doufexi, C.~Williams, and J.~McGeehan, ``Cognitive Node Selection
and Assignment Algorithms for Weighted Cooperative Sensing in Radar
Systems,'' IEEE Wireless Communications and Networking Conference, April 2009, pp. 1-6.

\bibitem{7500412_RadioMap}
F.~Paisana, Z.~Khan, J.~Lehtomäki, L.~A. DaSilva, and R.~Vuohtoniemi,
``Exploring Radio Environment Map Architectures for Spectrum Sharing in the
Radar Bands,'' 23rd International Conference on Telecommunications (ICT), May 2016, pp. 1-6.

\bibitem{7537114_MIMO_Com_5GH}
J.~Um, J.~Park, and S.~Park, ``Multi-Antenna-Based Transmission Strategy in 5
GHz Unlicensed Band,'' Eighth International Conference on
	Ubiquitous and Future Networks (ICUFN), July 2016, pp. 651-655.

\bibitem{rotatingRadar1}
R.~Saruthirathanaworakun, J.~Peha, and L.~Correia, ``{Opportunistic Sharing
	between Rotating Radar and Cellular}," \emph{IEEE Journal on Selected Areas in Communications}, \textbf{ 30}, 10, November 2012, pp. 1900-1910.

\bibitem{7131242_AirTrafficRadar}
H.~Wang, J.~Johnson, C.~Baker, L.~Ye, and C.~Zhang, ``On Spectrum Sharing
between Communications and Air Traffic Control Radar Systems,'' IEEE Radar Conference (RadarCon), May 2015, pp. 1545-1550.

\bibitem{5872307_Wimax}
A.~Lackpour, M.~Luddy, and J.~Winters, ``{Overview of Interference Mitigation
	Techniques Between WiMAX Networks and Ground Based Radar},'' in Annual Wireless and Optical Communications Conference (WOCC), April 2011, pp. 1-5.

\bibitem{1593335_haykin}
S.~Haykin, ``Cognitive radar: a way of the future,''  in \emph{IEEE Signal Processing Magazine,} \textbf{23}, 1, January 2006, pp. 30-40.

\bibitem{martone2014cognitive}
A.~Martone, ``Cognitive Radar Demystified,'' in \emph{URSI Bulletin}, \textbf{350}, September 2014, pp. 10-22.

\bibitem{6956678_CogRadar_OverviewSpecSending}
M.~T. Mushtaq, F.~A. Butt, and A.~Malik, ``An Overview of Spectrum Sensing in
Cognitive Radar Systems,'' IEEE Microwaves, Radar and Remote
	Sensing Symposium (MRRS), September 2014, pp. 115-118.

\bibitem{7410950_CogRadar_SpecSensing}
P.~Stinco, M.~S. Greco, and F.~Gini, ``Spectrum sensing and sharing for
cognitive radars,'' in \emph{IET Radar and Sonar Navigation,} \textbf{10}, 3, March 2016, pp. 595-602.

\bibitem{deng2013}
H.~Deng and B.~Himed, ``{Interference Mitigation Processing for
	Spectrum-Sharing between Radar and Wireless Communications Systems},'' in
\emph{{IEEE Transaction on Aerospace and Electronic Systems}}, in \emph{IEEE Transaction on Aerospace and Electronic Systems,} \textbf{49}, 3, July 2013, pp. 1911-1919.

\bibitem{6933960_AdaptiveRadarBeamforming}
Z.~Geng, H.~Deng, and B.~Himed, ``Adaptive Radar Beamforming for Interference
Mitigation in Radar-Wireless Spectrum Sharing,'' in \emph{IEEE Signal Processing Letters,} \textbf{22}, 4, April 2015, pp. 484-488.

\bibitem{6636787_Radar_Precoder}
A.~Babaei, W.~H. Tranter, and T.~Bose, ``A Practical Precoding Approach for
Radar/Communications Spectrum Sharing,'' International
	Conference on Cognitive Radio Oriented Wireless Networks, July 2013, pp. 13-18.

\bibitem{sodagari2012}
S.~Sodagari, A.~Khawar, T.~Clancy, and R.~McGwier, ``A projection Based
Approach for Radar and Telecommunication Systems Coexistence,'' IEEE
	GLOBECOM '12, December 2012, pp. 5010-5014.

\bibitem{6817773_SS_SpatialApprch}
A.~Khawar, A.~Abdel-Hadi, and T.~C. Clancy, ``{Spectrum Sharing between S-Band
	Radar and LTE Cellular System: A Spatial Approach},'' IEEE
	International Symposium on Dynamic Spectrum Access Networks (DYSPAN), April 2014, pp. 7-14.

\bibitem{6956861_ImpactOfChannel}
A.~Khawar, A.~Abdelhadi, and T.~C. Clancy, ``{On the Impact of Time-Varying
	Interference-Channel on the Spatial Approach of Spectrum Sharing Between
	S-band Radar and Communication System},'' IEEE Military Communications Conference (MILCOM), October 2014, pp. 807-812.

\bibitem{6810549_radarwaveformDesign}
K.~D. Shepherd and R.~A. Romero, ``Radar Waveform Design in Active
Communications Channel,'' Asilomar Conference on Signals,
	Systems and Computers, November 2013, pp. 1515-1519.

\bibitem{radarwaveAmuru}
S.~Amuru, R.~Buehrer, R.~Tandon, and S.~Sodagari, ``{MIMO Radar Waveform Design
	to Support Spectrum Sharing},'' IEEE Military Communications Conference (MILCOM), November 2013, pp. 1535-1540.

\bibitem{awais2014}
A.~Khawar, A.~Abdel-Hadi, and T.~Clancy, ``{MIMO Radar Waveform Design for
	Coexistence With Cellular Systems},'' IEEE International Symposium
	on Dynamic Spectrum Access Networks (DYSPAN), April 2014, pp. 20-26.

\bibitem{7472362_MutualInfo_Joint}
M.~Bica, K.~W. Huang, V.~Koivunen, and U.~Mitra, ``Mutual Information Based
Radar Waveform Design for Joint Radar and Cellular Communication Systems,''
IEEE International Conference on Acoustics, Speech and Signal
	Processing (ICASSP), March 2016, pp. 3671-3675.

\bibitem{bhat2012}
S.~Bhat, R.~Narayanan, and M.~Rangaswamy, ``Bandwidth Sharing and Scheduling
for Multimodal Radar with Communications and Tracking,'' IEEE 7th Sensor
	Array and Multichannel Signal Processing Workshop (SAM), June 2012, pp. 233-236.

\bibitem{martone3}
A.~Martone, K.~Sherbondy, K.~Ranney, and T.~Dogaru, ``Passive Sensing for
Adaptable Radar Bandwidth,'' IEEE Radar Conference (RadarCon), May 2015, pp. 0280-0285.

\bibitem{6586097_ofdmWaveform}
S.~Gogineni, M.~Rangaswamy, and A.~Nehorai, ``Multi-Modal OFDM Waveform
Design,'' IEEE Radar Conference (RadarCon), April 2013, pp. 1-5.

\bibitem{7131098_Joint}
J.~R. Guerci, R.~M. Guerci, A.~Lackpour, and D.~Moskowitz, ``Joint Design and
Operation of Shared Spectrum Access for Radar and Communications,'' IEEE Radar Conference (RadarCon), May 2015, pp. 0761-0766.

\bibitem{7470514_OptimumCoDesign}
B.~Li, A.~P. Petropulu, and W.~Trappe, ``Optimum Co-Design for Spectrum Sharing
between Matrix Completion Based MIMO Radars and a MIMO Communication
System,'' in \emph{IEEE Transactions on Signal Processing,} \textbf{64}, 17, September 2016, pp. 4562-4575.

\bibitem{7485158_MIMO_RadarClutter}
B.~Li and A.~Petropulu, ``MIMO Radar and Communication Spectrum Sharing with
Clutter Mitigation,'' IEEE Radar Conference (RadarCon), May 2016, pp. 1-6.

\bibitem{7472289_JointDesign}
B.~Li, H.~Kumar, and A.~P. Petropulu, ``A Joint Design Approach for Spectrum
Sharing between Radar and Communication Systems,'' IEEE International Conference on Acoustics, Speech and Signal Processing 	(ICASSP), March 2016, pp. 3306-3310.
	
\end{thebibliography}

\end{document}